\documentclass[12pt,twoside]{article}

\usepackage{amsfonts}
\date{} 

\begin{document} 

\centerline{\bf Adv. Studies Theor. Phys. 5(11) 545-549 2011} 

\centerline{} 

\centerline{} 

\centerline {\Large{\bf A note on Hardy's paradox}} 

\centerline{} 

%\centerline{\Large{\bf }} 

\centerline{} 

\centerline{\bf {J.F. Geurdes}} 

\centerline{} 

\centerline{C. vd. Lijnstraat 164 2593 NN Den Haag Netherlands} 

\centerline{han.geurdes@gmail.com} 

%\centerline{Address of Author1 third line} 

%\centerline{Address of Author1 forth line} 

\centerline{} 

\begin{abstract}A classical probabilistic explanation for Hardy's 'measurement after mutual annihilation' quantum paradox is demonstrated.
\end{abstract}

{\bf Keywords:} Classical probability, locality, Hardy paradox.

\section{Classical probability}
\label{ClassProb}
The common view on classical probability is that the triple, $(\Omega,\mathbf{F},P)$ cannot explain quantum effects like measurement after annihilation and tunneling. Here the sample space $\Omega$ can be {\it any} non-empty set. A $\sigma$-field, $\mathbf{F}$ is obtained from the set of all subsets, $\mathbf{P}(\Omega)=2^{\Omega}$, of $\Omega$. $\mathbf{F}$ is called a $\sigma$-field \cite{Ros} if, (i) $\Omega \in \mathbf{F}$, (ii) $E \in \mathbf{F} \Rightarrow E^{c}=(\Omega - E) \in \mathbf{F}$, (iii) $E,F,... \in \mathbf{F} \Rightarrow E \cup F \cup .... \in \mathbf{F}$.  The triple is completed with a probability measure $P$, such that, $\left(\forall:{X\in \mathbf{F}}\right) (0\leq P(X)\leq 1),~P(\Omega)=1$. 

\section{Pre-measurement characteristics}
Let us inspect the possibilities of a classical probability triple for Hardy's paradox \cite{Hardy}. In this paradox, quantum particles like electron and positron can be measured {\it after} mutual annihilation in a double Mach-Zehnder interferometer experiment. This possibility appears to reject the existence of pre-measurement characteristics \cite{Eins} and in this way seems to establish the completeness of a non-locality view on quantum paradoxes. We will study the possibilities of set systems designed to picture numerals  \cite{Jaquet} for a representation of physical events like particles being at a certain position in the double Mach-Zehnder experiment.
\\
\\
Apart from zero, unity and two, the numeral sets of von Neuman and of Zermelo are disjoint. This fact may represent mutual exclusion of electron and positron at the annihilation wing of the double Mach-Zehnder interferometer. The annihilation in experiment can be established by the escape of a photon. So the claim is that classical probability is unable to explain the $\frac{1}{16}$ probability from quantum mechanics that the photon is observed and shortly thereafter two separated detector clicks simultaneously occur. 

\section{ Numerals and algebra}
The numeral sets are defined subsequently. We have, $D_0=C_0=\emptyset$. Von Neuman numerals are $(n=0,1,2,3,...)$
\begin{equation}
C_{n+1} = \left\{C_0,C_1,.....,C_n\right\}.
\label{e9}
\end{equation}
Hence, $C_1=\left\{\emptyset\right\}$, $C_2=\left\{\emptyset,\left\{\emptyset\right\}\right\}$, $C_3=\left\{\emptyset,\left\{\emptyset\right\},\left\{\emptyset,\left\{\emptyset\right\}\right\}\right\}$, etc. 
\\
Zermelo's numeral system is 
\begin{equation}
D_{n+1} = \left\{D_n\right\}.
\label{e10}
\end{equation}
Hence, $D_1=\left\{\emptyset\right\}$, $D_2=\left\{\left\{\emptyset\right\}\right\}$, $D_3=\left\{\left\{\left\{\emptyset\right\}\right\}\right\}$, etc.
\\
\\
We establish mutual exclusion (annihilation) with sets $C(\hat{x} )$ and $D(\hat{x} )$ . Here $\hat{x}$ is $(x_1,x_2,x_3,x_4)$, with, $x_k \in \mathbb{R}$, and, $k=1,2,3,4$. We have
\begin{equation}
C(\hat{x}) = C_3(x_1)\cup C_3(x_2)\cup D_3(x_3)\cup D_3(x_4),
\label{e10a}
\end{equation}
together with
\begin{equation}
D(\hat{x}) = C_3(x_4)\cup C_3(x_3)\cup D_3(x_2)\cup D_3(x_1).
\label{e10b}
\end{equation}
Here, $x_k$ for $k=1,2,3,4$ substitutes the $\emptyset$ in (\ref{e9}) and (\ref{e10}). When, $x_1 \neq x_2$, $x_2 \neq x_3$, $x_3 \neq x_4$ and $x_4 \neq x_1$, randomly chosen from the real axis, it is easy to acknowledge that $C(\hat{x}) \cap D(\hat{x}) = \emptyset$. The sample space is defined by $\Omega = C(\hat{x}) \cup D(\hat{x})$. 
\\
\\
Because all constituent sets of $C(\hat{x})$ and $D(\hat{x})$ are mutually disjoint and the cardinality of $C_3(x_n)$, $n \in \{1,2,3,4\}$ denoted with, $|C_3(x_n)|$ equals 3 and $|D_3(x_m)|=1$, $m \in \{1,2,3,4\}$, we see that $ |\Omega | = 2\times(3+3+1+1)=16$.  This entails the $\sigma$-field $\mathbf{F}=\mathbf{P}(\Omega)=2^{\Omega}$. The set of all subsets of $\Omega$, has cardinality $|\mathbf{F}| = 2^{16}$. 
\\
\\
Given the structure of the numeral sets and their use in equations (\ref{e10a}) and (\ref{e10b}) we also may note that e.g. $C_2(x_n) \in \mathbf{F}$ and $D_2(x_m) \in \mathbf{F}$ and $\mathbf{F}$ is an algebra (see section \ref{ClassProb}). Finally, let us define the probability measure for $\Omega$ as $P(X)=|X|/|\Omega|= {1\over{16}}|X|$. Hence, we have $P\sim Uniform(\Omega)$ and $(\Omega,\mathbf{F},P)$ establishes a classical probability triple. 
\\
\\
Let us, subsequently, introduce the 'monadic union of a set' \cite{Union}, \cite{Hajnal} operation on a set $Z$, 
\begin{equation}
\cup [Z] = \left\{ x | (\exists : y \in Z) (x \in y) \right\}
\label{e13a}
\end{equation}
\\
Note that the pseudo-numeral sets in (\ref{e10a}) and (\ref{e10b}) refer to the mutual exclusion of two particles e.g. an electron and a positron in the annihilation wing of a Mach-Zehnder interferometer in Hardy's experiment. We suppose that the monadic union unleashes the hidden pre-measurement characteristics.

\section{Hardy's physics} 
In Hardy's paradox there is measurement of residuals after mutual annihilation. The claim is that classical probability cannot explain this phenomenon. However, let us suppose that $A(x_1)=C_3(x_1)$. The set $A(x_1)$ is supposed to be a hidden extra parameter of the $C(\hat{x})$ particle. We have, $A(x_1)\subset C(\hat{x})$. Similarly, $B(x_1)=D_3(x_1)$ and  $B(x_1) \subset D(\hat{x})$. Of course, $A(x_1) \cap B(x_1) =\emptyset$. 
\\
\\
Subsequently let us note that in the annihilation we are free to employ the monadic union of a set operator defined in (\ref{e13a}) and arrive at $\cup [A(x_1)] = C_2(x_1)$ and $\cup [B(x_1)] = D_2(x_1)$. Here, we remain within the $\mathbf{F}$ algebra. Now as we can observe from the definitions of von Neuman's and Zermelo's numerals, we have $C_2(x_1) \cap D_2(x_1) = D_2(x_1)$. Hence, with the probability measure defined previously, it is easy to arrive at 
\begin{equation}
P((\cup [A(x_1)])\cap (\cup [B(x_1)])) = {1\over{16}}.
\label{e13b}
\end{equation}
which is exactly the probability that quantum mechanics predicts for the measurement of two particles after mutual annihilation in the Hardy paradox \cite{Hardy}. Hence, when the monadic union of a set is allowed as a physical operation in the mutual annihilation, then a classical physics explanation is found for a typical quantum mechanical effect and it is no longer justified to reject extra hidden parameters basing oneself upon Hardy's experiment.

\section{Conclusion}
A classical probabilistics explanation for a typical quantum behavior, similar to tunneling, has been found. If the monadic $\cup$ operation as defined in (\ref{e13a}) cannot be excluded from physics it may represent a 'quantum' physical process and establishes a classical probabilistic explanation. It must be noted that implicitly there is also the question if sets of different type, $j_x$ and $j_y$, with, $j_y=j_x+1$, if, $x \in y$, can be allowed as representing particles. Of course a rejection of this as well as of the use of the monadic union in the annihilation {\it must} be based on hard physical principles. 
\\
\subsection{Physical picture}
A possible physical picture for $\cup$ can perhaps be associated to a 'dark' mirror-matter sector \cite{Foot}, \cite{Okun}, \cite{Footb} that may arise as a consequence of the experimentally established weak interaction parity non-invariance \cite{Lee}, \cite{Wu}. One can imagine that in a particle-wave duality sense, some unioned form of subset of electron and positron set representation survives the annihilation. Noting the fact that mirror sector and ordinary sector matter may only interact through gravity but noting too that Einstein's gravity equations can be transformed into Dirac's relativistic quantum equation \cite{JFG}, \cite{JFGx} a 'hiding in a mirror sector through gravity transformation' is a genuine physical picture for the previous described classical probability explanantion of Hardy's paradox.

\end{document}